%% file: main.tex
\documentclass{article}
\usepackage{PRIMEarxiv}

\usepackage[inline]{enumitem}
\usepackage{pbox}
\usepackage{balance}
\usepackage{todonotes}
\newlist{inlinelist}{enumerate*}{1}
\setlist*[inlinelist,1]{label=\roman*),itemjoin={{, }},itemjoin*={{, and }}}
\usepackage{url}
\usepackage[normalem]{ulem}
\usepackage{multirow}
\usepackage{multicol}
\usepackage{booktabs}
\usepackage{amsfonts}
\usepackage{amsmath}
\usepackage{amssymb}
\usepackage{amsthm}
\usepackage{adjustbox}
 \usepackage[normalem]{ulem}
 \useunder{\uline}{\ul}{}

\title{Naver Labs Europe (SPLADE) @ \\ TREC Deep Learning 2022}

\author{
     Carlos Lassance, Stephane Clinchant \\
  Naver Labs Europe \\
  France\\
  \texttt{carlos.lassance@naverlabs.com, stephane.clinchant@naverlabs.com}
}

\begin{document}

\maketitle

\begin{abstract}
This paper describes our participation to the 2022 TREC Deep Learning challenge. We submitted runs to all four tasks, with a focus on the full retrieval passage task. The strategy is almost the same as 2021, with first stage retrieval being based around SPLADE, with some added ensembling with ColBERTv2 and DocT5. We also use the same strategy of last year for the second stage, with an ensemble of re-rankers trained using hard negatives selected by SPLADE. Initial result analysis show that the strategy is still strong, but is still unclear to us what next steps should we take.

\end{abstract}

\section{Introduction}
\input{introduction}

\section{Methodology}

\input{method}

\section{Analysis on MS MARCO v1 and v2}
\input{msmarco}

\section{TREC DL 2022 - initial analysis}
\input{trec}

\section{Conclusion}
\input{conclusion}

\bibliographystyle{apalike}  
\bibliography{main}  

\appendix
\input{appendix}

\end{document}

%% file: introduction.tex
In this paper, we detail our TREC 2022 Deep Learning track submission, based on the latest improvements of the SPLADE model~\cite{efficiency,pp}. Compared to last year, there were only three changes: \begin{inlinelist} \item The use of Rocchio~\cite{rocchio} with SPLADE to improve first stage ranking \item A first stage ensemble of different SPLADE models, ColBERTv2~\cite{colbertv2} and DocT5~\cite{docT5} \item Inclusion of a MonoT5 based reranker using T0pp \end{inlinelist}. In total, we submitted more than 30 runs, most of them baselines using the models from~\cite{efficiency,pp} that are available at \url{https://huggingface.co/naver}. As per last year, we focus on passage retrieval and for the document task, we simply score a document by taking the maximum score over its passages.

This year we decided to make this notebook more streamlined compared to last year. For a more thorough introduction of the models used here, we invite the reader to check the following articles: SPLADE training~\cite{pp,efficiency}, ColBERTv2\cite{colbertv2}, DocT5~\cite{docT5}, Training style we used for our rerankers~\cite{gao2021rethink} and T5 based reranking~\cite{nogueira2020document}.

%% file: method.tex
In the following, we introduce the models we consider for both candidate generation as well as re-ranking. We also describe our training procedure, and detail the submitted runs.

\subsection{First Stage}

For the first stage this year, we only have one addition which is the use of Rocchio on SPLADE via Anserini. Notably: we use the following methods \begin{inlinelist} \item SPLADE++ EnsembleDistil~\cite{pp} \item SPLADE++ SelfDistil~\cite{pp} \item ColBERTv2\cite{colbertv2} \item DocT5~\cite{docT5} \end{inlinelist}

\subsection{Second Stage}

As per last year competition we use a mix of different PLMs as rerankers for which training is inspired by~\footnote{code available at \url{https://github.com/luyug/Reranker}}~\cite{gao2021rethink} and using negatives from SPLADE. Namely we use the following PLMs: Deberta-v2 xxlarge, Deberta-v3 large, Electra-large, T0pp, Albert-v2-xxlarge, Roberta-Large. We make the rerankers available at \url{https://huggingface.co/naver/\{name\}} with the models being named: trecdl22-crossencoder-\{debertav2,debertav3,electra,albert,roberta\}.

One main difference from last year is that we added to the mix a reranker based on MonoT5~\cite{nogueira2020document}. For that reranker we start from the T0pp model which is an 11B PLM. We train it both with the MonoT5 loss and the InfoLCE from~\cite{gao2021rethink}. Unfortunately, this model did not worked as well as it would be expected from its size, and in hindsight would be better served to go fully on the InfoLCE loss, as demonstrated by a post-TREC paper~\cite{https://doi.org/10.48550/arxiv.2210.10634}.

\subsection{Ensembling}

We also have applied ensembling in order to improve our results. This year we used ranx~\cite{bassani2022ranx} to generate all our ensembles, using average normalized score over the ensembles, unless explicitly noted. The normalized score uses the min and max values of the query so that, for each model, the best score is 1 and the lowest one 0.

\subsection{Runs submitted to TREC}

We submitted a total of 32 runs, 16 for the passage task and 16 for the document task. All of the document runs are the same as the passage ones, just with the added max pooling over the passages over the same document. For the 16 runs, we have the 3 official full ranking runs, 3 official rerank runs and 10 baselines runs (5 with rocchio, 5 without). 

\subsubsection{Official full ranking runs}

For our three full-ranking runs, we always use the same 6 rerankers for the second stage, while we vary the first stage by adding a new model for each new run:

\begin{itemize}
    \item \emph{\text{NLE\_SPLADE\_RR}}: First stage ensemble of \text{naver/splade-cocondenser-ensembledistil} and \text{naver/splade-cocondenser-selfdistil}, both models using Rocchio, followed by the ensemble reranking of 6 models.
    \item \emph{\text{NLE\_SPLADE\_CBERT\_RR}}: Same as before, but adding ColBERTv2~\cite{colbertv2}~\footnote{\url{https://downloads.cs.stanford.edu/nlp/data/colbert/colbertv2/colbertv2.0.tar.gz}} to the first stage ensemble.
    \item \emph{\text{NLE\_SPLADE\_CBERT\_DT5\_RR}}: Same as the previous one, but adding DocT5~\cite{docT5} to the first stage ensemble.
\end{itemize}

\subsubsection{Official rerank runs}

For our three official reranking runs, we submit the ensemble of the 6 rerankers using either the average normalized score or average normalized Condorcet score and a run based on T0pp.

\subsubsection{Baseline runs}

For our baseline runs, the idea was to simply run the four models from our SIGIR 2022 contributions~\cite{pp,efficiency} that have been made available in huggingface and an ensemble of the two best Splade++ models. To those 5 runs, we also add the Rocchio versions of each run, totaling 10 baselines.

%% file: msmarco.tex
In order to generate our final runs, we used as development scores the dev set of MSMARCO v1 and the last three TREC competitions. Results for our final runs are available in Table~\ref{tab:msmarco}. From it we drew some conclusions (passage track):

\begin{enumerate}
    \item Rocchio~\cite{rocchio} improves the performance of SPLADE when labels are not sparse (i.e. outside of MSMARCO v1 dev)
    \item As expected, TREC19, 20, and 21 are biased to techniques that participated in the competition, we notice that the ensemble of SPLADE models is way more competitive in 21 where it was first applied, while the effectiveness of DT5 greatly diminishes over time (as more techniques are added to the fold)
    \item Moreover, we are able to ``beat'' the best nDCG@10 results for TREC2019 and 21, but not for 2020, while we are able to increase mAP in all years. The TREC2020 reranker that got the best results seems very impressive, especially because it gets those results reranking BM25 directly.
    \item There is not so much correlation between results on the MSMARCO dev and TREC sets. For example, SPLADE\_EFFICIENT\_V performs the best out of all baselines on MSMARCO dev but is not on the TOP5 when we consider the average of TREC results.
    \item More-so, while our efficient models get very good results on MSMARCO, they struggle on TREC (something we already had seen). We are still not sure of the reason for this decrease in performance
    \item Statistically significant results are hard to get on TREC. We do not report on the Table, but most first-stage models are considered ``the same'' when we apply statistical significance testing, maybe there would be a way of joining all years in order to get more queries and thus more significant results? The average as we use here does not seem like a good idea (see point 1)
    \item Over the rerankers it seems that Electra-large took the cake as the best model we trained, while T0pp struggled on MSMARCO, but was surprisingly good on TREC21. However, adding the models on an ensemble almost always helped.
    \item While the rerankers improve over the first stage models, it is slightly deceiving as they add a lot more cost for inference. However, this is something that seems to be changing (the gap is larger on TREC21)
\end{enumerate}

\input{tables/msmarco.tex}

%% file: tables/msmarco.tex
\begin{table}[ht]
\centering
\label{tab:msmarco}
\caption{Results for the developmental process of our runs. *:SPLADE models considered without rocchio for ensembles evaluated on MSMARCO dev set.}
\adjustbox{max width=\textwidth}{%
\begin{tabular}{c|c|cc|cc|cc|cc}
\toprule
\multirow{2}{*}{Reranker} & Dev MSMARCO v1 & \multicolumn{2}{c|}{TREC 2019} & \multicolumn{2}{c|}{TREC 2020} & \multicolumn{2}{c|}{TREC 2021} & \multicolumn{2}{c}{Average over TREC} \\
                                      & MRR@10 & nDCG@10 & mAP@1000 & nDCG@10 & mAP@1000 & nDCG@10 & mAP@100 & nDCG@10 & mAP  \\ \midrule
BASELINE - Best@year                  & 46.3   & 76.5    & 50.49      & 80.3    & 56.43      & 74.9    & 39.23     & 77.2    & 48.7 \\ \midrule
\multicolumn{10}{c}{Runs sent as baselines}                                                                                         \\ \midrule
SPLADE\_PP\_SDISTIL                   & 37.8   & 73.6    & 50         & 72.8    & 51.4       & 68      & 32.4      & 71.5    & 44.6 \\
SPLADE\_PP\_EDISTIL                   & 38.3   & 73.2    & 50.5       & 72.0    & 50         & 68.5    & 33.3      & 71.2    & 44.6 \\
SPLADE\_EFFICIENT\_V                  & \textbf{38.8}   & 71.5    & 47.4       & 71.5    & 48.4       & 67.9    & 31.9      & 70.3    & 42.6 \\
SPLADE\_EFFICIENT\_VI-BT              & 38.0   & 70.3    & 46.2       & 69.8    & 47.4       & 65.7    & 29.3      & 68.6    & 41.0 \\
SPLADE\_PP\_SDISTIL\_ROCCHIO          & 32.9   & 71.2    & 51.4       & 72.8    & 50.8       & 70.0    & 34.6      & 71.3    & 45.6 \\
SPLADE\_PP\_EDISTIL\_ROCCHIO          & 32.8   & 71.5    & 50.7       & 73.7    & 53.0       & 69.7    & 34.4      & 71.6    & 46.0 \\
SPLADE\_EFFICIENT\_V\_ROCCHIO         & 32.2   & 70.1    & 49.1       & 72.6    & 49.8       & 67.7    & 33.0      & 70.1    & 44.0 \\
SPLADE\_EFFICIENT\_VI-BT\_ROCCHIO     & 30.2   & 71.4    & 49.5       & 67.2    & 46.1       & 64.1    & 29.0      & 67.6    & 41.5 \\
SPLADE\_ENSEMBLE\_PP                  & 38.4   & 72.7    & 50.4       & 73.1    & 51.4       & 69.2    & 33.9      & 71.7    & 45.2 \\
SPLADE\_ENSEMBLE\_PP\_ROCCHIO         & 33.7   & 71.8    & 51.2       & 73.7    & 52.5       & 70.8    & 35.3      & 72.1    & 46.3 \\ \midrule
\multicolumn{10}{c}{First Stage ensembles (not sent. just a base of comparison)}                                                             \\ \midrule
Run 1 - NLE\_SPLADE                   & 38.4*   & 71.8    & 51.2       & 73.7    & 52.5       & 70.8    & 35.3      & 72.1    & 46.3 \\
Run 2 - NLE\_SPLADE\_CBERT          & \textbf{39.5*}   & 72.8    & 52         & 75.7    & 54.2       & 71.5    & 36.9      & 73.3    & 47.7 \\
Run 3 - NLE\_SPLADE\_CBERT\_DT5     & 39.3*   & \textbf{76.3}    & \textbf{54.3}       & 75.3    & 54.3       & 70.7    & 37.1      & 74.1    & 48.6 \\ \midrule
\multicolumn{10}{c}{Rerankers over Run 3 (not sent. just a base of comparison)}                                                             \\ \midrule

Deberta-v2 xxlarge & 41.5 & 75.7          & 56.0          & 76.6 & 57   & 73.6 & 39.9          & 75.3 & 51.0 \\
Deberta-v3 large   & 40.7 & \textbf{77.3} & \textbf{57.2} & 75.7 & 57   & 73.8 & 39.5          & 75.6 & 51.2 \\
Electra-large & \textbf{41.9} & 76.9 & \textbf{57.3} & \textbf{77.1} & \textbf{57.6} & \textbf{74.2} & \textbf{39.9} & \textbf{76.1} & \textbf{51.6} \\
T0pp               & 38.8 & 73.5          & 55.1          & 74.5 & 55.4 & 73.4 & \textbf{39.9} & 73.8 & 50.1 \\
Albert-v2-xxlarge  & 41.7 & 77.2          & 57.0          & 76.9 & 57.3 & 73.4 & 38.5          & 75.8 & 50.9 \\
Roberta-Large      & 41.5 & 75.3          & 55.6          & 75.8 & 56.2 & 69.8 & 36.5          & 73.6 & 49.4 \\ \midrule

\multicolumn{10}{c}{Final runs}                                                                                                       \\ \midrule
Run 1 - NLE\_SPLADE\_RR               & \textbf{43.1}   & 78.6    & 58.8       & \textbf{79.2}    & \textbf{59.9}       & 75.5    & 42.5      & 77.8    & 53.7 \\
Run 2 - NLE\_SPLADE\_COLBERT\_RR      & 43.0   & 78.4    & 59.0       & 79      & 59.3       & 75.4    & 42.6      & 77.6    & 53.6 \\
Run 3 - NLE\_SPLADE\_COLBERT\_DT5\_RR & 42.9   & \textbf{78.8}    & \textbf{59.9}       & 78.7    & 59.4       & 75.5    & 42.7      & 77.7    & 54.0 \\ \bottomrule
\end{tabular}
}
\end{table}

%% file: trec.tex
As per the previous year, we mostly focused on the passage track, and thus report only results and analysis for it. Results are made available on Table~\ref{tab:trec} . We drew some initial conclusions and are still analyzing results:

\begin{enumerate}
    \item The gap between the first stage and reranked runs increased compared to previous years.
    \item The gap between our best run and the best possible run was stable compared to last year (0.102 difference last year, 0.106 this year), however our best run increased the gap to the ``median'' run\footnote{the run that got the median score on all queries, which is different from the ``median'' submission} (0.135 last year, and 0.177 this year). While the former shows that we maintained the goodness of our runs, the latter is probably a consequence of point 1 (larger gap between first stage and reranking).
    \item Our three runs performed almost the same, which is expected given the results we had seen on the development set, but is kinda frustrating as improving the first stage ranking seems to plateau after reranking.
    \item As it had happened on the development set, Rocchio improved the results from SPLADE.
    \item There seems to be a lack of consistency on our ensembling process, making that various of our runs got ``the best'' result for different queries. What is more worrying is the case of query 2002533, where our first stage models got the best nDCG@10 over all submitted runs, but our rerankers got a more average score. More intelligent ensembling seems to be needed. More details on Table~\ref{tab:queries}
    \item Over all our trained networks we only used the training set of MSMARCO v1. 
    \item Complementing the last point, there were not so many new things we added this year, and currently, it seems that we would do the same for next year as well.
\end{enumerate}

\input{tables/trec.tex}

\begin{table}[ht]
\centering
\caption{Some query examples and comments}
\label{tab:queries}
\begin{tabular}{c|p{2cm}|p{4cm}|p{4cm}}
    \toprule
    query id & query & overall results & our results \\ \midrule
    2053884 & when a house goes into foreclosure what happens to items on the premises & Has two distinct wh propositions (when and what). It was the query with the highest distance between median nDCG@10 (0.07) and best nDCG@10 (1.0) & SPLADE and BM25 are completely lost in this query (0 nDCG@10), but rerankers saved the day. \\ 
    2002533 & how much average cost to plan a 8' tree? & model has to understand that 8' is 8 feet. Has the worst best nDCG@10 at 0.42 and median is pretty low as well (0.07) & Differently from the previous one, here our first stage models shine (best nDCG@10 at 0.42) and the rerankers suffered (0.24) \\
    2003157 & how to cook frozen ham steak on nuwave oven & not so sure why it was so hard, maybe nuwave got badly tokenized? Had the worst PREC@10 (best model got only 3 positives in the top10). & Our models perform well, very close to the best nDCG@10 and we get the best PREC@10. \\
    2028378 & when is trial by jury used & also not sure why it is so hard, but had the worst best mAP (only 0.06). It has too many positives and many duplicates & Average results for our models, but the number of duplicates makes our mAP suffer\\ \bottomrule
\end{tabular}
\end{table}

%% file: tables/trec.tex
\begin{table}[ht]
\centering
\label{tab:trec}
\caption{Results for the TREC 2022 passage track.}
\adjustbox{max width=\textwidth}{%
\begin{tabular}{c|cc}
\toprule
Method                                             & ndcg@10 & mAP (100) \\ \midrule
\multicolumn{3}{c}{Runs sent as baselines}                                          \\ \midrule
SPLADE\_PP\_SDISTIL                                & 57.05    & 18.46      \\
SPLADE\_PP\_EDISTIL                                & 57.86    & 18.01      \\
SPLADE\_EFFICIENT\_V                               & 55.09    & 16.31      \\
SPLADE\_EFFICIENT\_VI-BT                           & 52.71    & 14.52      \\
SPLADE\_PP\_SDISTIL\_ROCCHIO                       & 58.97    & 19.68      \\
SPLADE\_PP\_EDISTIL\_ROCCHIO                       & 59.17    & 19.23      \\
SPLADE\_EFFICIENT\_V\_ROCCHIO                      & 54.52    & 17.25      \\
SPLADE\_EFFICIENT\_VI-BT\_ROCCHIO                  & 50.84    & 14.52      \\
SPLADE\_ENSEMBLE\_PP                               & 57.89    & 18.62      \\
SPLADE\_ENSEMBLE\_PP\_ROCCHIO                      & 59.91    & 20.05      \\ \midrule
\multicolumn{3}{c}{Runs}                                                     \\ \midrule
Run 1 - NLE\_SPLADE\_RR                            & 70.92    & 29.77      \\
Run 2 - NLE\_SPLADE\_COLBERT\_RR                   & 71.41    & 29.63      \\
Run 3 - NLE\_SPLADE\_COLBERT\_DT5\_RR              & 71.45    & 29.50     \\ \bottomrule
\end{tabular}
}
\end{table}

%% file: conclusion.tex
For the TREC DL 22 competition, we submitted runs based on our recent SPLADE advancements for first-stage ranking, followed by an ensemble of re-rankers trained with hard negatives selected by SPLADE. While this is a very rough draft, and we are still analyzing the results, it seems that this year there is a larger gap between the first stage runs and the reranked ones. Also, compared to the best results possible per query, our submitted runs get results that are inline with last year.